\documentclass[12pt]{iopart}
\usepackage{iopams}
\usepackage{setstack}
\usepackage[english]{babel}
\usepackage{cite}
\usepackage[pdftex]{graphicx}
\usepackage[pdftex]{hyperref}
\usepackage{lscape} 
\usepackage{pdflscape} 
\usepackage{rotating}
\usepackage[usenames]{color}
\usepackage[normalem]{ulem}
\usepackage{ulem} 
\usepackage{subcaption}

\newcommand{\abs}[1]{\left| #1 \right|} 
\newcommand\bluesout{\bgroup\markoverwith{\textcolor{blue}{\rule[0.5ex]{4pt}{1.5pt}}}\ULon} 

\begin{document}

\title[]{Informational analysis of the confinement of an electron in an asymmetric double quantum dot\footnote{\textbf{Accepted for publication in Physica B: Condensed Matter (2024).\href{https://doi.org/10.1016/j.physb.2024.416769}{{\color{blue} Click here to access.}}}}}

\author{W. S. Nascimento$^{1,\dagger}$, A. M. Maniero$^{2,\ddagger}$, F. V. Prudente$^{1,\S}$, C. R. de Carvalho$^{3,\P}$, Ginette Jalbert$^{3,**}$}

\address{$^1$Instituto de F\'{\i}sica, Universidade Federal da Bahia, 40170-115, Salvador, BA, Brazil. 

$^2$Centro das Ci\^encias Exatas e das Tecnologias, Universidade Federal do Oeste da Bahia, 47808-021, Barreiras, BA, Brazil. 

$^3$Instituto de F\'{\i}sica, Universidade Federal do Rio de Janeiro, 21941-972, Rio de Janeiro, RJ, Brazil.}

\ead{$^{\dagger}$wallassantos@gmail.com, $^{\ddagger}$angelo.maniero@ufob.edu.br, $^{\S}$prudente@ufba.br, $^{\P}$crenato@if.ufrj.br, $^{**}$ginette@if.ufrj.br.}

\vspace{10pt}


\begin{abstract}
A{\it quasi-}unidimensional one-electron double quantum dot is studied within the framework of Shannon informational entropy. Its confinement potential, which is described by an asymmetric harmonic-gaussian function, consists of two wells separated by a potential barrier, and the asymmetry of the potential in respect to the center of the barrier is parameterized by $x_0$. In this work we employ Shannon informational entropy as a tool to investigate changes in the electronic confinement resulting from the modifications in the degree of asymmetry $x_0$. In particular, we notice that Shannon entropy turns out to be very sensitive to the change from the symmetric to the antisymmetric regime. Moreover, we show that Shannon entropy as a function of $x_0$ for an electronic excited state displays an irregular behavior whenever the state energy is in the vicinity of a local extreme (a maximum or a minimum); connected with this we observe an oscillatory behavior of the energy of the excited states as a function of $x_0$. \\

\end{abstract}

%
\vspace{1pc}

%
%
%
%

\section{Introduction}

Confined quantum systems form a class of problems in physics that have received attention since their pioneering studies in the mid-1930s~\cite{michael,darwin}. Nowadays, with the impressive increase in computer processing capacity and the improvements in algorithms designed to describe the confinement of electrons, nuclei, atoms or molecules, research in this area has once again gained great relevance~\cite{Juan-Jose_etal2023,Coomar_etal2022, Yanajara-Parra_etal2024,Deshmukh_etal2021}. In this context, quantum dots have been the subject of several researches, mainly due to the fact that these nanostructures have numerous technological applications, for example, in the manufacture of transistors~\cite{kastner1992,lee-etal:2023}, solar cells~\cite{Wu-Wang2014,PhysRevApplied.13.044035} and LEDs-diodes~\cite{Jang2023,Yu2023}.

Quantum dots (QDs) are nanometer-long heterostructures consisted of semiconductor materials~\cite{chiquito_2001,Bimberg-etal1999}. Such structures correspond to confinement potentials with well-defined energy levels for electrons that can be deployed one by one starting from zero. Consequently QDs can be modeled as artificial atoms where electrons are confined in the spatial directions $\widehat{x}$, $\widehat{y}$ and $\widehat{z}$~\cite{Ashoori1996,kastner2000,Sako-Diercksen2003}. On the other hand, from two coupled neighboring QDs one obtains what is called a double QD, which can work as an artificial molecule.

Studies involving single or double QDs cover from the study of the electron's transport properties in a single few-electron QD~\cite{Kouwenhoven-etal2001} or the determination of the absolute number of electrons in each dot of a double few-electron QD~\cite{Chan-etal2004} to the computation of macroscopic physical quantities \cite{mandal2015,Sargsian2023,ccakir2017linear}. Calculations involving the electronic structure of these systems are performed at different levels of accuracy using the Hartree approximation~\cite{pfannkuche1993, creffield2000}, the Hartree-Fock computation~\cite{thompson2005, olavo2016} or the full conﬁguration interaction method~\cite {thompson2005, olavo2016}, among others. Furthermore, different functions have been proposed to describe the QD's confinement potential such as harmonic potential functions~\cite{maniero2020a, maniero2021}, inverted gaussian~\cite{Adamowski2000,maniero2023}, polynomial ~\cite{Mukherjee_etal2015, Mukherjee2016} or harmonic-gaussian~\cite{Cheng-Wang2019,Qiao_etal2023}. More precisely, the double-well harmonic-gaussian potential function, as defined in this work and in Refs.~\cite{Nascimento_etal2024,duque-2023-magnetico,duque2023-laser}, has been used in the composition of the confinement potential function of the symmetric and asymmetric double QD. 

The Shannon entropy was originally formulated in the informational context, more precisely, in studies involving sending and receiving of messages in a communication system ~\cite{shannon1948a,shannon1948b}. Within the framework of information theory applied to quantum-mechanical, we define the Shannon informational entropies in the space of positions, $S_r$, and in momentum space, $S_p$, beyond the entropic sum, $S_t$~\cite{sen2011,nascimento-prudente2018}. Additionally, we witness the emergence of different informational quantities such as Fisher information~\cite{Fisher_1925,Estanon_etal2021} and Kullback–Leibler entropy~\cite{Kullback-Leibler1951,Majtey_etal2005}. In this regard, physical systems such as harmonic oscillator ~\cite{Olendski2023,nascimentocoulomb-et_al2021}, hydrogen atoms~\cite{nascimento-prudente2018,jiao-et_al2017}, helium~\cite{nascimentocoulomb-et_al2021,Anupam2023}, molecular systems~\cite{Onyeaju_etal2023,Nalewajski2006}, between others~\cite{Santos2022,wallas_fred_educacional,Esquivel_etal2023,Estanon_etal2023}, has been treated successfully from a quantum-informational point of view. 

Recently, Shannon entropy began to be used in works involving QDs, for example, in the analysis  of impurity in the InxGa1-xN QD~\cite{liu2023} and of the structural measures and entanglement in He atom in a QD~\cite{PhysRevA.105.032821}. In addition, we already have informational studies involving the influence of the magnetic field on circular dots~\cite{Shafeekali-Olendski2023} and the effect of temperature of the hydrogenic impurity state in GaAs quantum well~\cite{De-hua_etal2024}. In our previous work we carried out an informational study of electronic confinement in symmetric double QD consisting of a GaAs/AlGaAs heterostructure and we used Shannon informational entropy as a tool to map the degeneracy of quantum states and study the level of coupling/decoupling between neighboring QDs~\cite{Nascimento_etal2024}.

The objective of the present work is to analyze the electronic confinement in an asymmetric double QD formed by a GaAs/AlGaAs heterostructure through informational entropies. To this end, we describe the confinement potential by an asymmetric double-well harmonic-gaussian function along the $\widehat{x}$ direction, and two harmonic functions for the $\widehat{y}$ and $\widehat{z}$ directions. The asymmetry of the potential is therefore parameterized allowing us to use Shannon entropy as a tool to investigate the impact on the system due to the change in the symmetry condition of the potential. In particular, we observe an oscillatory behavior of the energy of the excited states as a function of the asymmetry parameter, and also an interesting connection between their extremes and the jumps in the values of Shannon entropy.

The present article is divided as follows: in Section~\ref{theo_form} we present the theoretical description of the system of interest and define the Shannon informational entropies; in Section~\ref{cal-met}  we discuss our calculation methodology where we point out the methods used to obtain the physical quantities of interest; in Section~\ref{ana and disc} we present our results and analysis; and, finally, in Section~\ref{conclusion} we summarize the main aspects of our investigation.     

\section{Theoretical formulation}\label{theo_form}

In this section, we present the theoretical formalism of our work. Here, we discuss the system of interest (subsection~\ref{systems_of_interest}) and the entropic quantities used in our study (subsection~\ref{information_entropies}). In this section atomic units (a.u.) are used.

\subsection{System of interest}\label{systems_of_interest}

A single electron confined in an asymmetric double QD formed by a GaAs/AlGaAs heterostructure can be described by the Hamiltonian
\begin{eqnarray}
H =- \frac{1}{2m_c}\nabla^2+ V(x,y,z) \ ,
\label{hamiltoniano}
\end{eqnarray}
where $m_c$ is the electron's effective mass and the potential confinement function $V(x,y,z)$ contains three terms, i.e.,
\begin{eqnarray}
V(x,y,z) = {V}_{ADQD}(x) + {V}_{HO}(y) + {V}_{HO}(z) \ ,
\label{potencialc}
\end{eqnarray}
where ${V}_{ADQD}(x)$ represents an asymmetric double-well harmonic-gaussian potential, and ${V}_{HO}(y)$ and ${V}_{HO}(z)$ are harmonic potentials. So the expression of ${V}_{ADQD}(x)$ is given by
\begin{eqnarray}
{V}_{ADQD}(x) =  V_0 \left[   A_1 \frac{x^2}{k^2} + A_2 e^{-\left(\frac{x}{k}-x_0\right)^2}  \right]  
\label{HGA}
\end{eqnarray}
whereas ${V}_{HO}(y)$ and ${V}_{HO}(z)$ are written as
\begin{eqnarray}
{V}_{HO}(y) = \frac{1}{2}m_c \omega^2_y y^2  \hspace{0.5cm}\mbox{and}\hspace{0.5cm}  {V}_{HO}(z) = \frac{1}{2}m_c \omega^2_z z^2 \ .
\label{HOYZ}
\end{eqnarray}
In Eq.~(\ref{HGA}) $x_0$ represents the asymmetry parameter,  $V_0$ is the depth of the well, and $k$ relates the well to the width of the barrier, influencing the confinement condition. Additionally, $A_1$ is associated with the well’s width, while $A_2$ corresponds to the height of the internal barrier and, consequently, the coupling between the two neighboring wells. The symmetric double-well harmonic-gaussian potential function discussed in our previous work~~\cite{Nascimento_etal2024} is a specific case of the function given by Eq.~(\ref{HGA}) when $x_0$~=0. Besides, in Eq.~(\ref{HOYZ}) the angular frequencies $\omega_y$ and $\omega_z$ are the confinement parameters along the  $\widehat{y}$ and $\widehat{z}$ directions.

\begin{figure*}[h] 
\includegraphics[scale=0.35]{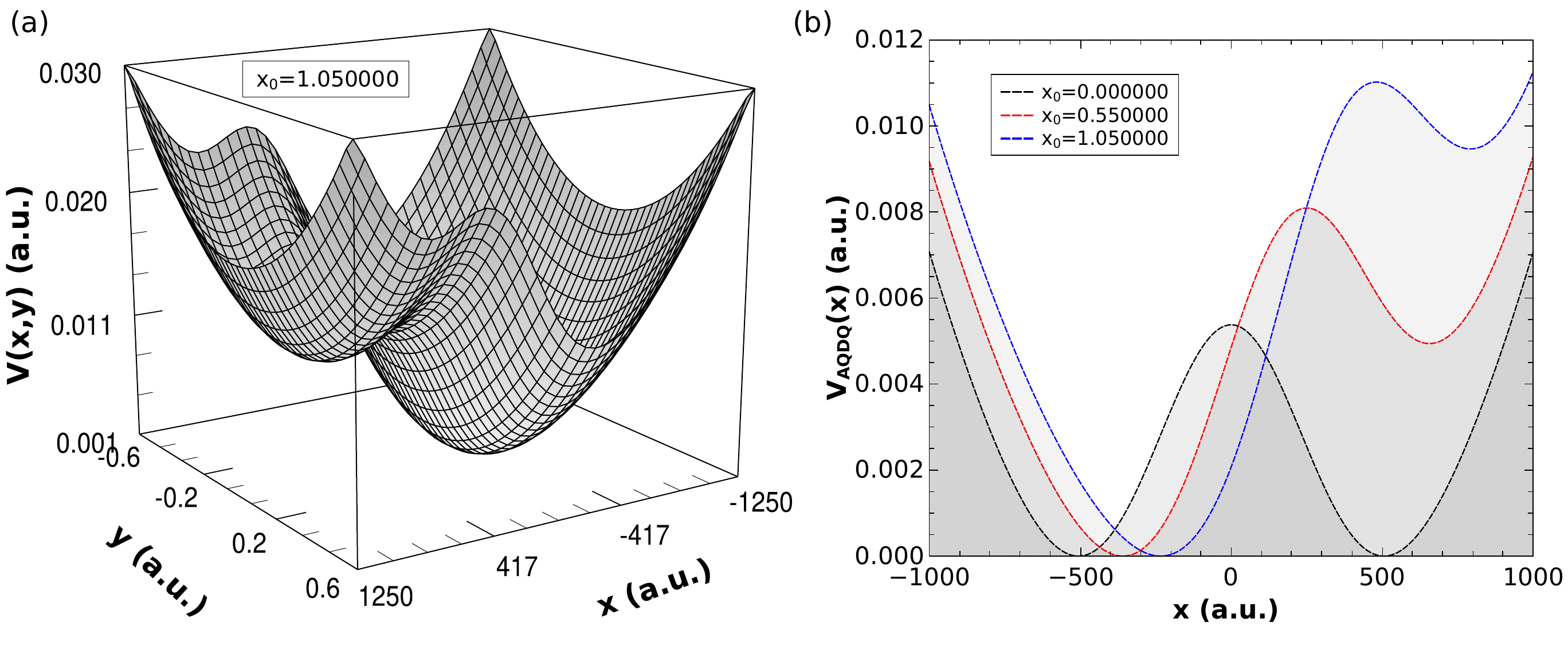}
\caption{(a)~Profile of the confinement potential ${V}(x,y,0)$. (b)~Profile of the asymmetric double-well harmonic-gaussian potential ${V}_{ADQD}(x)$ for various values of $x_0$. We used the following fixed parameters: $A_1$~=~0.2, $A_2$~=~1.2, $k$~=~377.945~a.u., $V_0$~=~0.00838~a.u., $\omega_y$~=~1.00~a.u e $m_c$~=~0.067~a.u..}
\label{figurainicial}
\end{figure*}

We study here the influence of asymmetry between the two wells of a double QD on the properties of the system. To this end, we  have fixed  the confinement conditions in the ${y}$ and ${z}$ directions. In this sense, in Fig.~\ref{figurainicial}a we observe the profile of $V(x,y,0)$, where we can perceive the asymmetry between the two adjacent wells of the double QD in respect to the internal barrier, and in Fig.~\ref{figurainicial}b we present potential profiles of ${V}_{ADQD}(x)$ for different values of $x_0$. Note that it is precisely the profile in the $\hat{x}$ direction that models the double well.

\subsection{Shannon informational entropies}{\label{information_entropies}}
In the framework of atomic and molecular physics, Shannon informational entropies in the space of positions, denoted as $S_r$, and in momentum space, denoted as $S_p$, are defined as ~\cite{sen2011,nascimento-prudente2018}
\begin{equation}
S_r = -\int \rho(x,y,z) \ln \left( \rho(x,y,z) \right) d^3r
\label{entropia_posicao}
\end{equation}
and
\begin{equation}
S_p = - \int \gamma(p_x, p_y, p_z) \ln \left( \gamma(p_x, p_y, p_z) \right) d^3p  \ ,
\label{entropia_momento}
\end{equation}
where the probability densities $\rho(x,y,z)$ and $\gamma(p_x, p_y, p_z)$ are normalized to unity and defined as: $\rho(x,y,z)=\abs{ \psi(x, y, z)}^2$ and $\gamma(p_x, p_y, p_z)=\abs{\widetilde{\psi}(p_x, p_y, p_z)}^2$. The momentum space wave function $\widetilde{\psi}(p_x, p_y, p_z)$ is obtained by applying a Fourier transform to the position space wave function $\psi(x, y, z)$. The entropies $S_r$ and $S_p$ are dimensionless quantities from the point of view of physics (for more details see Refs.~\cite{nascimento2018,nascimento-et_al2020,matta-etal2011}).

The quantities $S_r$ and $S_p$ serve as measures of delocalization or dispersion of the probability densities $\rho(x,y,z)$ and $\gamma(p_x, p_y, p_z)$, respectively~\cite{nascimento-et_al2020,Cruz2024}. It should be noted that informational entropies are calculated without taking into account a reference point in probability distribution, as it happens in the computation of standard deviation, which measures the average deviation from the mean value of the physical quantity involved, such as position, for example~\cite{nascimento-et_al2020}. Basically, the difference between the standard deviation in position and $S_r$ is that, if the density $\rho(x,y,z)$ has a profile with two narrow peaks that are far from each other, the standard deviation will be large while $S_r$ will be small. On the other hand, if the density profile consists of only one narrow (large) peak, both the standard deviation and $S_r$ will be small (large). In our previous work, Ref.~\cite{nascimento-et_al2020}, we inspected in more detail the standard deviation and information entropies as measures of quantum uncertainty.

The entropy sum, $S_t$, is defined by adding the entropies $S_r$ and $S_p$ which, in turn, originate the entropic uncertainty principle which reads~\cite{birula-mycielski1975,nascimento-prudente2018}
\begin{eqnarray}
S_t&=&S_r+S_p \nonumber \\
&=& - \int \int  \rho(x,y,z)  \ \gamma(p_x, p_y, p_z) \ \ln \left( \ \rho(x,y,z)  \ \gamma(p_x, p_y, p_z) \ \right) d^3r \ d^3p  \nonumber \\ 
&\geq & 3(1+\ln\pi) \label{St2} .
\end{eqnarray}
From the entropic uncertainty relation we can derive the Kennard uncertainty relation~\cite{birula-mycielski1975}. The quantity $S_t$ can be used to analyze the correlation between the densities $\rho(x,y,z)$ and $\gamma(p_x, p_y, p_z)$~\cite{kumar-prasad2023}.

\section{Calculation Methodology}\label{cal-met}  
We are interested in solving the Schr\"odinger equation
\begin{eqnarray}
H \psi(x,y,z) = E\psi(x,y,z) \ ,
\label{eq_sch}
\end{eqnarray}
where the Hamiltonian $H$ is given by Eqs.(\ref{hamiltoniano}--\ref{HOYZ}) and the one-electron wave function $\psi(x,y,z)$ is written in general form as
\begin{eqnarray}
\psi(x,y,z) &=&\sum^n_{n_x,n_y,n_z=0}\sum^2_{i=1}\sum_{\mu_i}  C^{n_x,n_y,n_z}_{i,\mu_i}\psi^x_{n_x,i,\mu_i}(x)   \psi^y_{n_y}(y)\psi^z_{n_z}(z). 
\label{funcao_geral}
\end{eqnarray}
The components $\psi^x_{n_x,i,\mu_i}(x)$, $\psi^y_{n_y}(y)$ and $\psi^z_{n_z}(z)$ are expressed in terms of base functions of the cartesian anisotropic gaussian orbital type as follows:  
\begin{eqnarray}
\psi^x_{n_x,i,\mu_i}(x) &=&  N_{n_x,i,\mu_i} (x-X_{_i})^{n_x}\exp[-\alpha^x_{\mu_i}(x-X_{i})^2]; \label{funcaox} \\
\psi^y_{n_y}(y) &=& N_{n_y} \ y^{n_y}\exp[-\alpha_y y^2] ;~\mbox{and} \label{funcaoy}\\
\psi^z_{n_z}(z) &=& N_{n_z} \ z^{n_z}\exp[-\alpha_z z^2] , \label{funcaoz}
\end{eqnarray}
where $N_{n_x,i,\mu_i}$, $N_{n_y}$ and $N_{n_z}$ are the normalization constants and the integers $n_x$, $n_y$ and $n_z$ allow the classification of the orbitals , for instance, $n = n_x + n_y + n_z=$ 0, 1, 2, ...., correspond to the types $s-$, $p-$, $d-$, respectively. We represent the function $\psi^x_{n_x,i,\mu_i}(x)$ as a gaussian function centered on $X_{i}$ ($i$ varying from 1 to 2). The $X_{i}$ represent the minima of $V_{{ADQD}}(x)$ and $C^{n_x,n_y,n_z}_{i,\mu_i}$, the expansion coefficients determined from the variational principle. The functions $\psi^y_{n_y}(y)$ and $\psi^z_{n_z}(z)$ are described in terms of a gaussian type function centered at the origin. 

In the present work, with respect to the general form of the one-electron wave function, Eqs.~(\ref{funcao_geral})-(\ref{funcaoz}), we employ few restrictions or simplifications: first, in Eq.~(\ref{funcaox}) we use two different gaussian exponents denoted as $\alpha^x_{\mu_i}$~--~the first exponent corresponds to a harmonic approximation within the respective well, and the second is $\frac{3}{2}$ times the value of the first one (for more details see Refs.~\cite{olavo2016, maniero2019,maniero2020a,maniero2021,maniero2023}); second, in Eqs.~(\ref{funcao_geral}),~(\ref{funcaoy})~and~(\ref{funcaoz}) we set $n_y=n_z=0$ to prevent excitations associated with the  $\widehat{y}$ and $\widehat {z}$ directions, whose energy scales are much higher than the one associated with the $\widehat {x}$ direction as we are considering large values of  $\omega_y$ and $\omega_z$ in order to approach a 1D problem; and finally  we set $\alpha_y=\alpha_z=\frac{1}{2}m_c\omega$ and $N_{n_y}=N_{n_z}=N$. It is worth mentioning that by setting $n_y=n_z=0$ the one-electron wave function $\psi(x,y,z)$ becomes separable in the Cartesian coordinates: $\psi(x,y,z) = \Phi_{n}(x)\psi^y_{0}(y)\psi^z_{0}(z)$.

We employ computational algorithms based on the variational method and matrix diagonalization techniques to solve the Schr\"odinger equation~Eq.(\ref{eq_sch}) and, thus, obtain the energy of the different eigenstates of the system. Additionally, the corresponding eigenfunctions are used to compute the Shannon entropies; in Ref.~\cite{Nascimento_etal2024} we made explicit the general forms of the densities $\rho(x,y,z)$ and $\gamma(p_x, p_y, p_z)$ and the corresponding Shannon informational entropies.

\section{Analysis and Discussion}\label{ana and disc}

In this section, we present and discuss our results regarding the energy contribution relative to the $\widehat{x}$ direction, $E_x^n$, and entropic quantities $S_r$, $S_p$ and $S_t$ as functions of $x_0$ for the values of $A_2=0.5$~and~1.2. 

The optimized wave function was expanded into the following basis
functions: on the $\hat{x}$ axis we employ orbitals of the type 2s2p2d2f2g1h (in total 11 functions located in each well) and on the $\hat{y}$ and $\hat{z}$ axes, 1s type orbitals.

We analyze the states of the first six energy levels and, for $A_2= 0.5$, we have two energy levels well below the central barrier, two other levels very close to it - one a little below and the other a little above it - and the last two levels well above it; whereas, for $A_2= 1.2$ , all energy levels lie well below the barrier. 

We disregard the energy contributions relative to the $\widehat{y}$ and $\widehat{z}$ directions, namely $E_y^0$ and $E_z^0$, in the analysis that follows, because they are not affected by the modifications in $x_0$ and  $A_2$. Besides, as $E_x^n \ll E_y^0, E_z^0$, we can only notice the variations of $E^n$ in a chart if we disregard $E_y^0$ and $E_z^0$.

In turn, the separability in Cartesian coordinates of $\psi(x,y,z)$ results in $S_r = S_x + S_y + S_z$ \cite{Nascimento_etal2024} and, hence, one should expect $S_y$ and $S_z$ to be disregarded on the same basis as $E_y^0$ and $E_z^0$. In fact, $S_y$ and $S_z$ are constants; they are not affected by the modifications in $x_0$ and  $A_2$. In this case, the variation of $S_x$ can be perfectly noticed in a graph if we plot $S_r$.

\subsection{On the energy and probability density of the states}{\label{energia1}}

\begin{figure}[h] 
\includegraphics[scale=0.36]{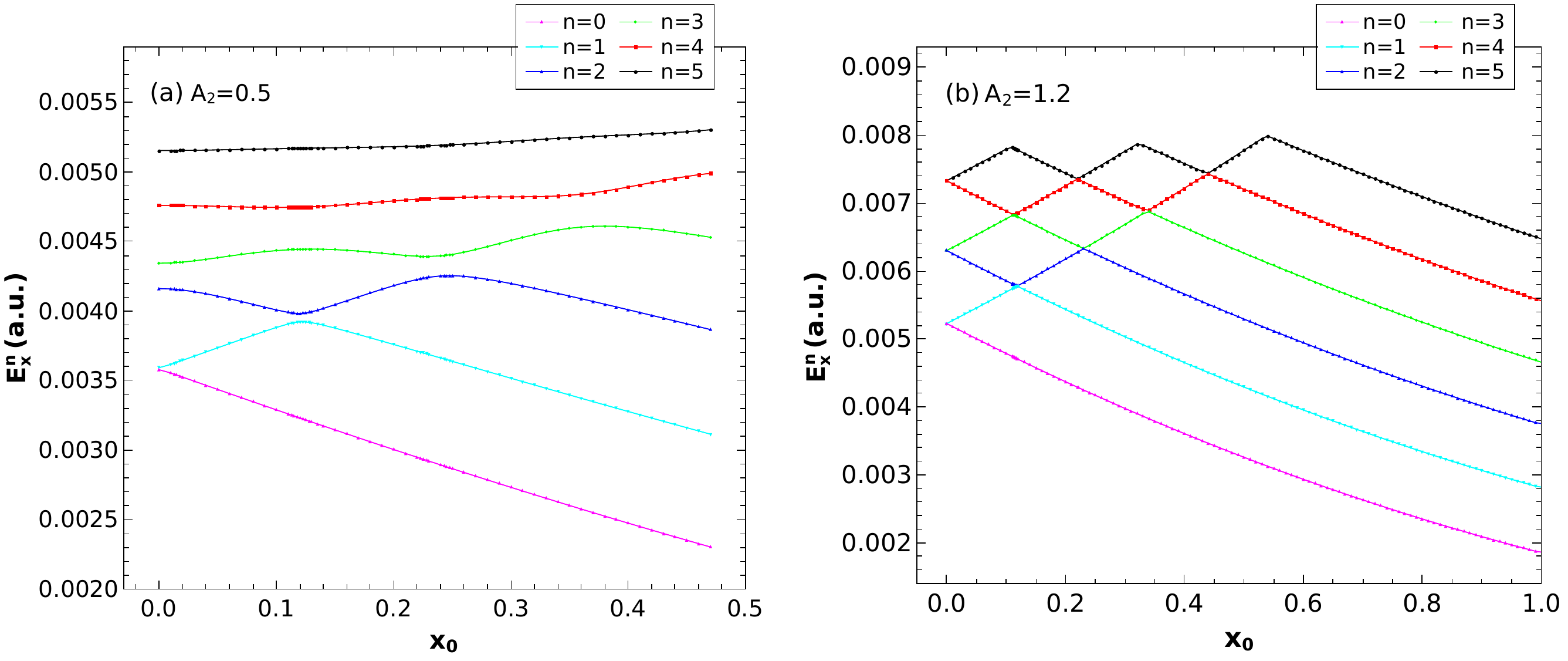}
\caption{$E_x^n$ as a function of  $x_0$, for  $n=0-5$. In \textbf{(a)} $A_2$~=~0.5 and \textbf{(b)} $A_2$~=~1.2.} 
\label{energia}
\end{figure}

First, let us comment the behavior of $E_x^n$ for the first six lowest states $n=0, 1,...5$ as function of the asymmetry parameter $x_0$. According to Fig.~\ref{energia}(a), where $A_2=0.5$, we observe that $E_x^{0}~\lesssim~E_x^{1}~<~E_x^{2}~<~E_x^{3}~<~E_x^{4}~<~E_x^{5}$; being the states $n=0$ and $n=1$ almost degenerate. In fact, according to Table~S1 of the supplementary material, one has $E_x^0 \approx 0.003578$ and $E_x^1 \approx 0.003593$. $E_x^n$ reveals a behavior pattern which becomes more evident in Fig.~\ref{energia}(b), where $A_2=1.2$: as long as the state's energy level is below or close to the central barrier, we notice that the higher $n$, the more persistent the oscillation of $E_x^n$ is, before assuming a decreasing behavior. This pattern is valid for the states $n \ge 1$; in the case of the state $n=0$, it always decreases with the increasing of $x_0$. Besides, we also observe that the deeper the energy levels are below the barrier, the more the pairwise formation of degenerate states occurs (see also Table~S2 of the supplementary material).

In a recent work, we reported an oscillatory behavior of some physical quantities of a two-electron QD as function of the magnetic field strength~\cite{maniero2020a, maniero2021}. In that case the oscillatory behavior is connected with the change in the ordering of the system's states on the energy scale. In the present case this does not happen, the order of the states on the energy scale remains the same. 

\begin{figure}[h] 
\includegraphics[scale=0.37]{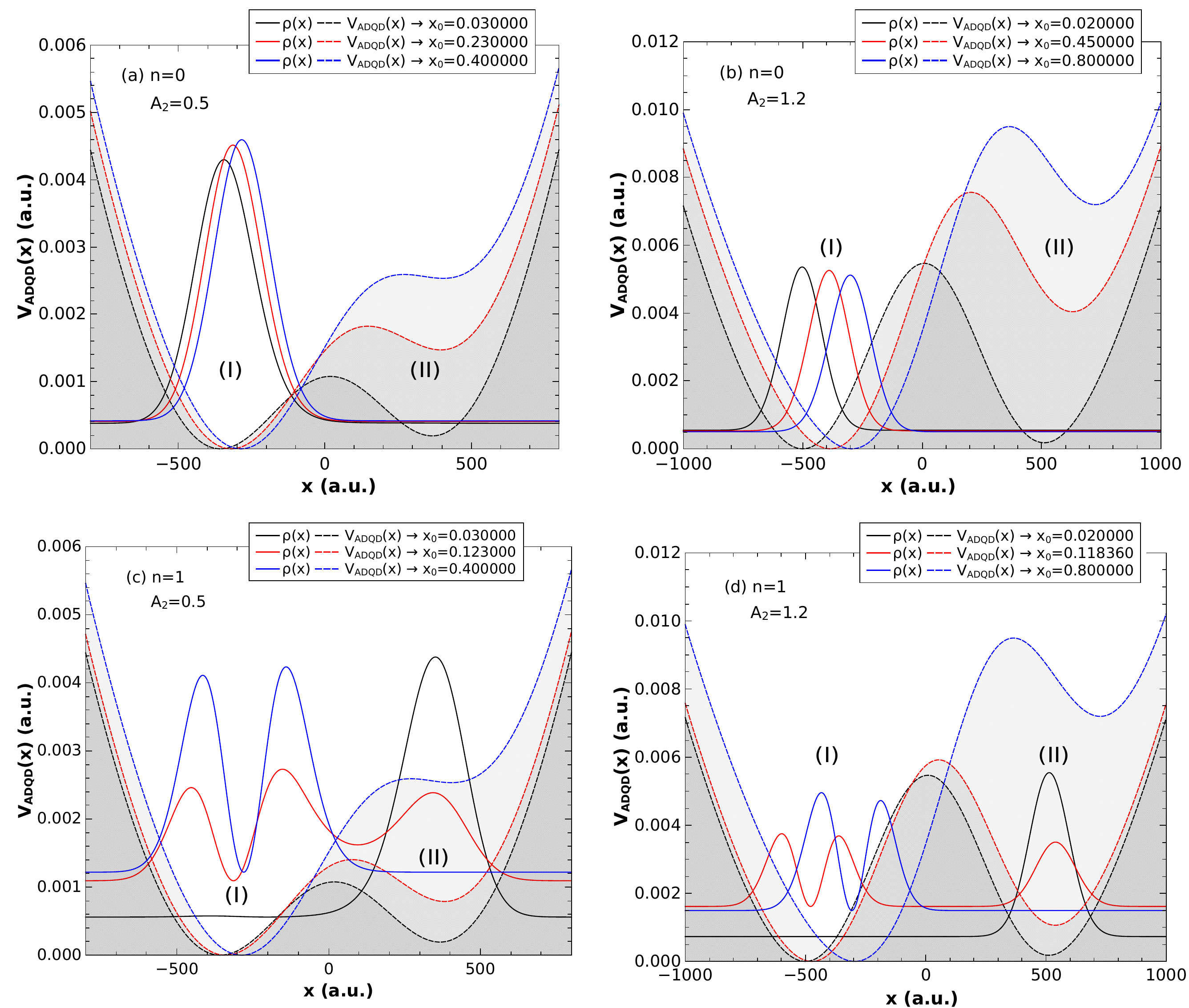}
\caption{Asymmetric double-well harmonic-gaussian potential, ${V}_{ADQD}(x)$, for three different values of $x_0$ together with the corresponding $\rho_0(x)$ and $\rho_1(x)$ when $A_2$~=~0.5~and~1.2. The shaded regions depict the profiles of ${V}_{ADQD}(x)$.}  
\label{potencialdensidade}
\end{figure}

The present oscillatory behavior is connected with the geometrical modification of the potential profile along the $\widehat{x}$ direction as the asymmetry parameter $x_0$ increases. The modifications in the confinement potential profile caused by changes in $x_0$, besides altering energy values, can lead to electron migration from one well to another of the potential. For example, in Fig.~\ref{potencialdensidade}(a)~and~(b), for $n$~=~0, the electron remains located in well I as $x_0$ increases. According to Fig.~\ref{potencialdensidade}(c)~and~(d), for $n$~=~1, the  electron migration occurs, that is, the electron, initially located necessarily in well II, begins to have a probability of being found in both wells, and finally only in well I as $x_0$ increases. Let us note that for $A_2$~=~0.5, Fig.~\ref{energia}(a), we have two energy levels well below the central barrier, two other levels very close to it - one a little below and the other a little above it - and the last two levels well above it; whereas, for $A_2$~=~1.2, Fig.~\ref{energia}(b), all energy levels lie well below the barrier. In the supplementary material we present a more detailed energy analysis. 

For the sake of completeness, in Fig.~S1 of the supplementary material is shown the curves of the probability densities in the moment space, $\gamma_x(p_x)$, corresponding to the configurations of the densities $\rho_0(x)$ and $\rho_1(x)$ of Fig.~\ref{potencialdensidade}.

\subsection{On the informational analysis}{\label{Informational analysis}}

\begin{figure}[h] 
\includegraphics[scale=0.36]{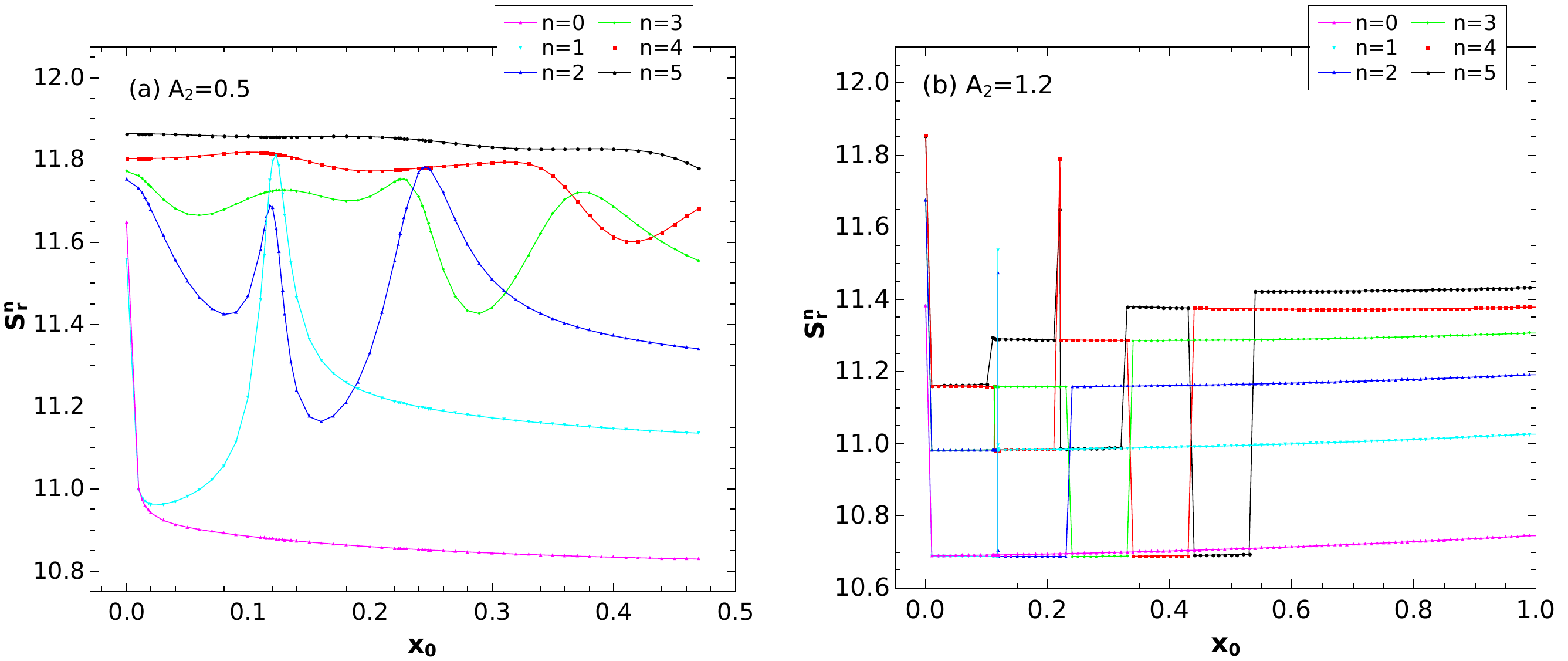}
\caption{$S_r^n$ as a function of $x_0$, for states $n=0-5$. In \textbf{(a)} $A_2$~=~0.5 and \textbf{(b)} $A_2$~=~1.2.} 
\label{Sr}
\end{figure}

In graphs (a) and (b) of Fig.~\ref{Sr} we present the Shannon entropy in the space of positions, $S_r^n$, as a function of the asymmetry parameter $x_0$, for states with $n$ ranging from 0 to 5 with $A_2~=~0.5$~and~1.2, respectively. The Tables~S3~and~S4 of the supplementary material contain all the computed values of $S_r^n$ obtained in this study. 

According to Fig.~\ref{Sr}(a), where $A_2$~=~0.5, both $S_r^0$ and $S_r^1$ vary significantly from $x_0$~=~0~to~0.01, resulting in a sharp drop in their curves at the very beginning of the graph on the left. In turn, the same behaviour is not observed for the entropies $S_r^{2-5}$, whose curves have a smooth variation for small values of $x_0$. On the other hand, when $A_2$~=~1.2, the abrupt behaviour of $S_r^0$ and $S_r^1$ for small values of $x_0$ ($0 < x_0 \ll 1$) observed in the previous case occurs for all entropies $S_r^{0-5}$, as we can see in Fig.~\ref{Sr}(b). All these behaviors of $S_r^{n}$ reflect the behaviors of the corresponding probability densities $\rho_n(x)$ and their degree of delocalization which are displayed in Fig.~S2 of the supplementary material. In the supplementary material we discuss in more detail the behavior of $\rho_n(x)$ when the symmetry condition is abandoned by changing sightly the asymmetry parameter $x_0$, from $x_0~=~0.00$ to $x_0~=~0.01$. In fact, a conclusion that can be reached is that $S_r^{n}$ are very sensitive to regime change, from symmetric to asymmetric, for those state whose energies are well below the central barrier separating the two wells of the double QD. 

We observe from Figs.~\ref{energia} and~\ref{Sr} or from the corresponding data displayed in tables S1 -- S4 in the supplementary material, is that $S_r^{n}$ displays a singular behavior, with large variation of its value, whenever the state $n$ has its degeneracy with another state removed or restored; even when degeneracy does not occur, we observe peaks in the $S_r^{n}$ curve when the extremes (maximum and minimum) of the corresponding $E_x^{n}$ gets close to the energy of a neighboring state.

For the sake of completeness, let us discuss the simple case of the state $n$~=~1. In the case of $A_2$~=~0.5, when the symmetric regime is broken by the introduction of a tiny asymmetry ($x_0 =0.0 \rightarrow 0.01$) $S_r^1$ evolves to a local minimum ($S_r^{1(min)}=10.96257180$) at $x_0=0.03$, point from which $S_r^1$ reaches a local maximum ($S_r^{1(max)}=11.81185483$) at $x_0^{max}=0.123$ and then decreases monotonically afterwards, see Fig.~\ref{Sr}(a). In the case of Fig.~\ref{Sr}(b), where $A_2=1.2$, after breaking the symmetry regime the increase of $x_0$ leads $S_r^1$ to a local minimum ($S_r^{1(min)}=10.68785746$) at $x_0=0.118$, afterwards to a local maximum ($S_r^{1(max)}=11.53684462$) at $x_0^{max}=0.11836$, point  from which $S_r^1$ decreases again. The maxima of $S_r^{1(max)}$ at $x_0^{max}=0.123$ and $0.11836$ correspond to the maxima degree of delocalization of $\rho_1(x)$ in Fig.~\ref{potencialdensidade}(c)~and~(d). A similar conclusion is reached for the other cases, $n=2 - 5$. In addition, in accordance with what is said in the previous paragraph, the maxima of $S_r^{1(max)}$ and $E_x^{1(max)}$ occur for the same value of $x_0^{max}$ for each specific value of $A_2$ according to Table~\ref{tabela}.

\begin{table}[h]
\center
\caption{Values of $S_r^{1{(max)}}$ and $E_x^{1{(max)}}$ with their respective values of $x_0^{max}$ for $A_2 = 0.5$ and $1.2$.}
\begin{tabular}{ccc}
\multicolumn{1}{l}{}       & \multicolumn{1}{l}{}             & \multicolumn{1}{l}{}             \\ \hline
\multicolumn{1}{|c|}{}     & \multicolumn{2}{c|}{\textbf{$A_2$}}                                              \\ \cline{2-3} 
\multicolumn{1}{|c|}{}     & \multicolumn{1}{c|}{0.5}        & \multicolumn{1}{c|}{1.2}        \\ \hline
\multicolumn{1}{|c|}{$S_r^{1{(max)}}$}   & \multicolumn{1}{c|}{11.81185483} & \multicolumn{1}{c|}{11.53684462} \\ \hline
\multicolumn{1}{|c|}{$E_x^{1{(max)}}$}    & \multicolumn{1}{c|}{0.00392196}  & \multicolumn{1}{c|}{0.00577788}  \\ \hline
\multicolumn{1}{|c|}{$x_0^{max}$} & \multicolumn{1}{c|}{0.123}    & \multicolumn{1}{c|}{0.11836} \\ \hline
\end{tabular}
\label{tabela}
\end{table}

\begin{figure}[h] 
\includegraphics[scale=0.36]{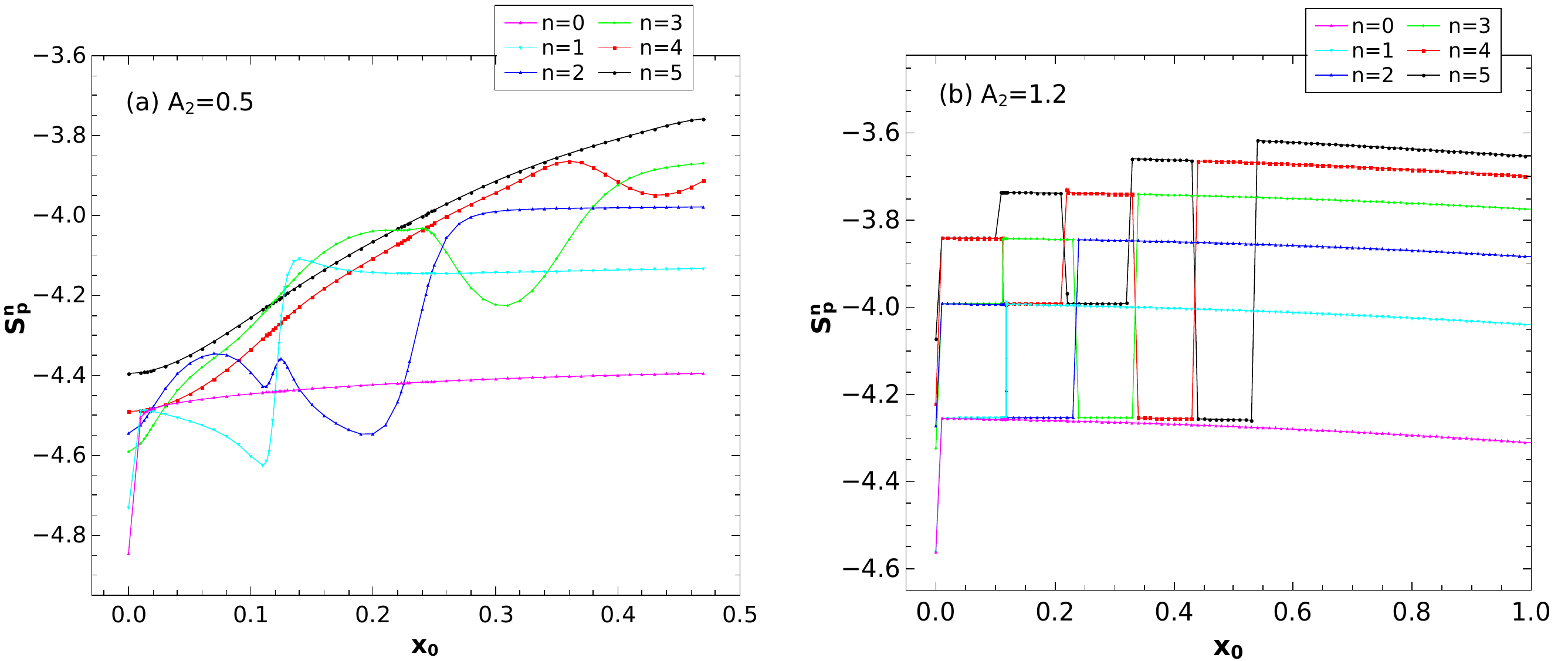}
\caption{$S_p^n$ as a function of $x_0$, for states $n=0-5$. In \textbf{(a)} $A_2$~=~0.5 and \textbf{(b)} $A_2$~=~1.2.} 
\label{Sp}
\end{figure}

In graphs (a) ($A_2$~=~0.5) and (b) ($A_2$~=~1.2) of Fig.~\ref{Sp}, we present the values of Shannon entropy in momentum space, $S_p^n$, as a function of $x_0$ for states $n=0-5$. In Tables S5 and S6 of supplementary material, we provide the values of $S_p^n$ obtained in this study.

\begin{figure}[h] 
\includegraphics[scale=0.36]{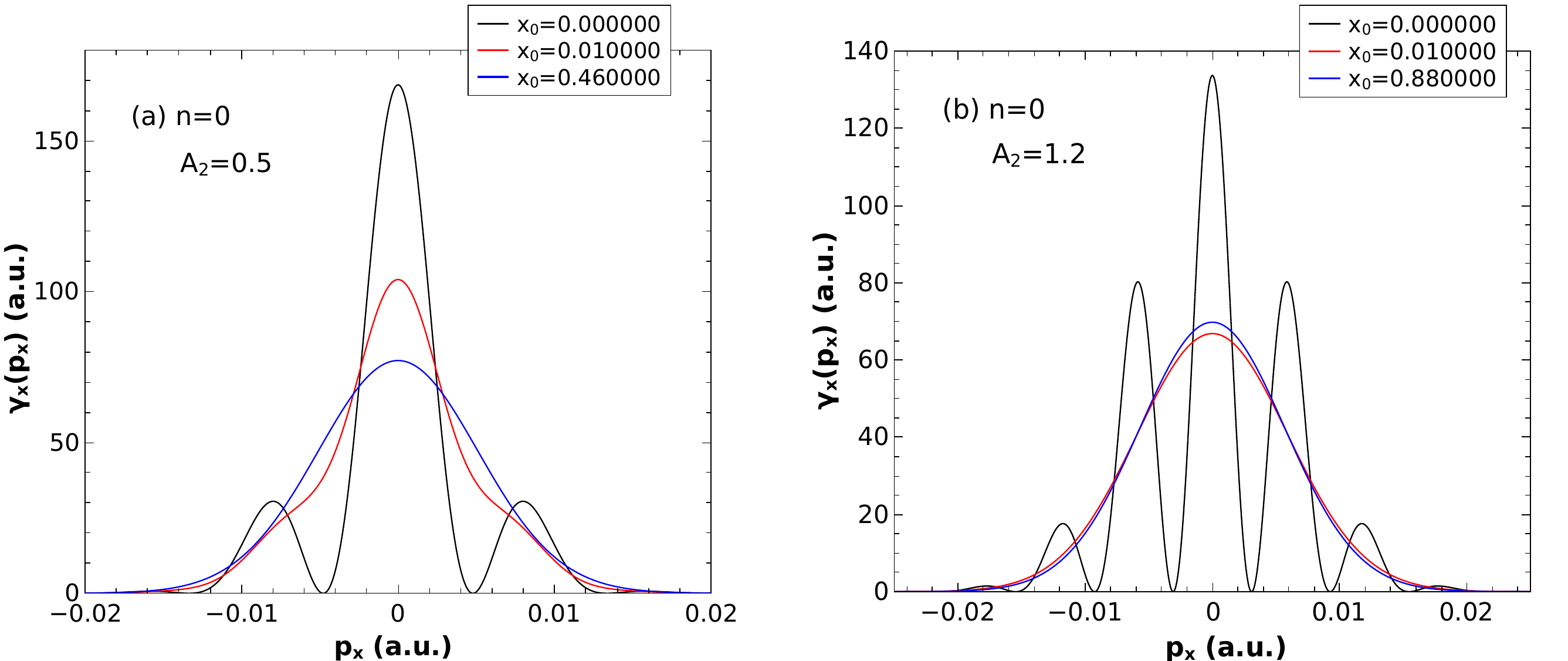}
\caption{$\gamma_x(p_x)$ as a function of $p_x$ for $n$~=~0. In \textbf{(a)} $A_2$~=~0.5 and \textbf{(b)} for $A_2$~=~1.2.} 
\label{densidadepx}
\end{figure}

In Fig.~\ref{Sp}(a) ($A_2$~=~0.5), it is shown that increasing $x_0$ leads to an increase of $S_p^{0}$. This effect is connected with the Fig.~\ref{densidadepx}(a), where we observe the increased of the degrees of delocalization in $\gamma_x(p_x)$ curves for $x_0$~=~0.00,~0.01~and~0.46. On the other hand, in Fig.~\ref{Sp}(b) ($A_2$~=~1.2) the values of $S_p^{0}$ increase when $x_0$ passes from 0.00 for 0.01 and decreases from $x_0=0.01$ onwards. Such behaviors are in agreement, for example, with the degrees of delocalization in $\gamma_x(p_x)$ curves for $x_0$~=~0.00,~0.01~and~0.88 in Fig. Fig.~\ref{densidadepx}(b). Similar analyses can be made for $S_p^{1-5}$.

In the quantum mechanical context, the informational entropies can indeed be negative as we can see in Tables~S5~and~S6 of the supplementary material. The Shannon's original work in the field of communication systems also suggests the possibility of obtaining negative values for informational entropy when working with continuous probability distributions \cite{shannon1948a}.

\begin{figure}[h] 
\includegraphics[scale=0.36]{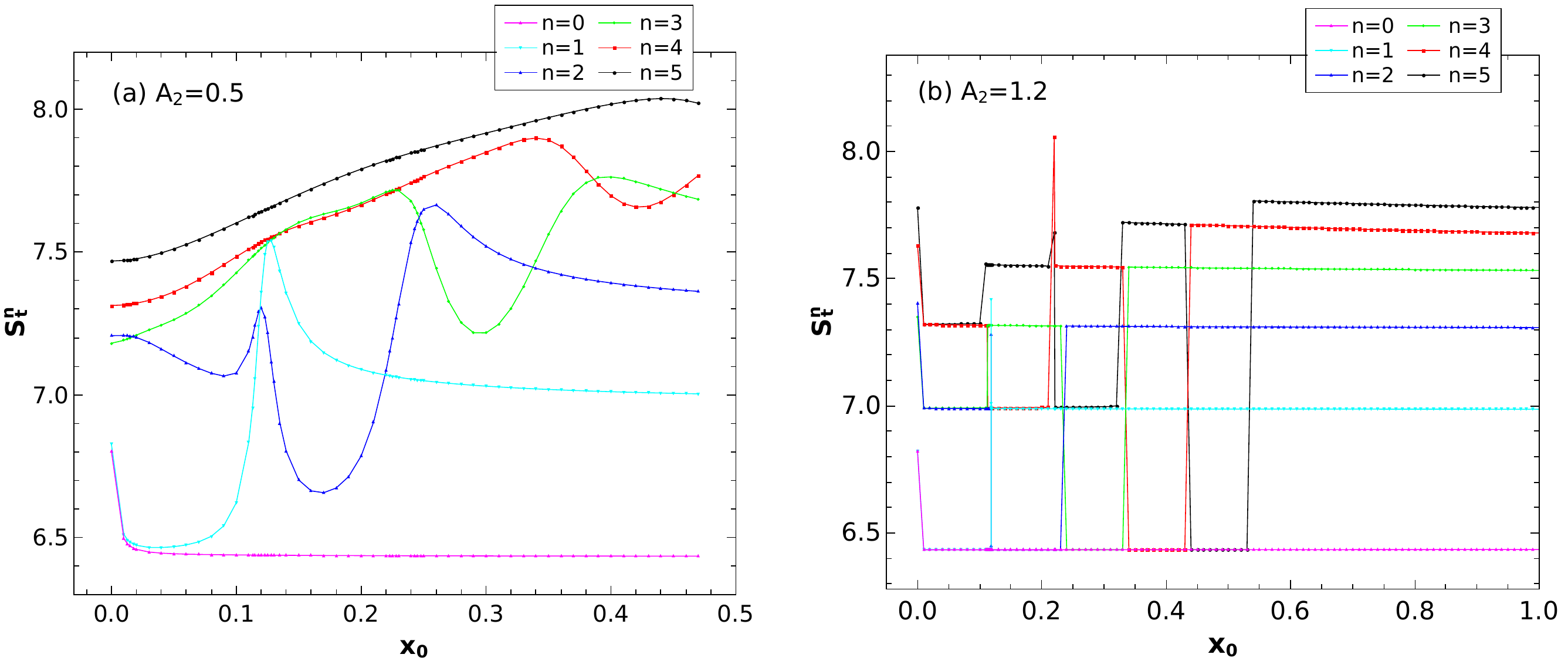}
\caption{$S_t^n$ as a function of $x_0$, for $n=0-5$. In \textbf{(a)} $A_2$~=~0.5 and \textbf{(b)} for $A_2$~=~1.2.} 
\label{St}
\end{figure}

In Figs.~\ref{St}(a)~e~(b), we observe that the profiles of the $S_t^n$ curves versus $x_0$ are similar to the profiles of $S_p^{n}$ from Fig.~\ref{Sp}(a)~and~(b). However, note that the maximum or minimum values of $S_t^{0-5}$ and $S_p^{0-5}$ do not necessarily coincide for the same values of~ $x_0$. 

We have shown in Ref.\cite{nascimento-et_al2020} that the value of $S_t^n$ for the harmonic oscillator is constant, regardless of mass and $\omega$, depending only on the quantum number. In turn, the present confinement potential $V_{ADQD}(x)$ (Eq.~\ref{HGA}) becomes harmonic for large $x_0$ and, consequently, in theory the value of the total entropy becomes approximately three times the value of $S_t$ for a one-dimensional harmonic oscillator. At least for the ground state this proposition is true. That is, the values of the total entropy for $x_0$~=~0.470000 ($A_1$~=~0.5) and $x_0$~=~1.910000 ($A_1$~=~1.2) are respectively, $S_t^{0}$~=~6.43453369~and~6.43555392. In turn, the value of three times $S_t^{0}$ of the one-dimensional harmonic oscillator is 6.4341.  

The entropic uncertainty relation~(\ref{St2}) is respected throughout the work. The inequality~(\ref{St2}) indicates that the variations in the values of $S_r$ and $S_p$ must occur in such a way that preserves a minimum value of 6.43418966. In this sense, both entropies $S_r$ and $S_p$ can even increase, for example, in the range of $x_0$ values between 0.110000 and 0.123000 (see Tables~S3~and~S5). 

\section{Conclusion}\label{conclusion}

We investigated the influence of asymmetry level on the properties of a system with an electron confined in an asymmetric double QD. We adopted a confinement potential, ${V}(x,y,z)$, defined by an asymmetric harmonic-gaussian potential function, ${V}_{ADQD}(x)$, in addition to harmonic potential functions ${V}_{HO}(y)$ and ${V}_{HO}(z)$. Although this is a 3D problem, we reduced it to a 1D problem, assuming that no excitations associated with the $\widehat{y}$ and $\widehat{z}$ directions can be achieved, as we considered ${V}_{HO}(y)$ and ${V}_{HO}(z)$ to be confinement potentials of large strength. Hence, we studied only the energy contribution along the $x$ axis, $E_x^{n}$, as well as the behavior of entropies $S_r$, $S_p$ and $S_t$ as a function of the asymmetry parameter $x_0$.
 
We identified in the symmetric regime, characterized by the asymmetry parameter $x_0=0.0$, the occurrence of degeneracy between the states whose energies are below the central barrier that separates the two wells of the double QD. For the case where all states satisfy this condition, what happens when $A_2=1.2$, the formation of the degeneracy occurs pairwise. For the two analyzed values of $A_2$, $0.5$ and $1.2$, the curves of $E_x^{0}$  decrease as $x_0$ increases, while the curves of $E_x^{1-5}$ exhibit oscillations as $x_0$ grows. Connected with this, we discussed the  electronic spacial behavior through the density curves $\rho_n(x)$.

We observed that $S_r^{n}$ is a quantity very sensitive to the regime change from symmetric to asymmetric, when the energy of the corresponding state $E_x^n$ is well below the central barrier. Moreover, we noticed that $S_r^n$ reveals a singular behavior which is in connection with the removal or restoration of  degeneracy between state $n$ and a neighboring one. And, even when there is no degeneracy involved, $S_r^n$ presents peaks in its curve whenever $E_x^n$ displays a minimum or a maximum getting close to the energy of a neighboring state.

Finally, we computed the values of $S_p^{n}$ and $S_t^{n}$ for $n=0-5$ when $A_2$ is 0.5 and 1.2. Regarding the Shannon entropy in momentum space, we explained the oscillations in the curves of $S_p^{0}$ versus $x_0$ by considering the delocalization of the density $\gamma_x(p_x)$ for $n=0$. Analogous analyses can be applied to interpret the curves of $S_p^{1-5}$ versus $x_0$. Overall, we discussed the profiles of the curves of $S_t^{0-5}$ versus $x_0$, and we find that the entropic uncertainty relation holds consistently throughout our study. \\ \\

\noindent \textbf{Supporting Information}

This manuscript contains supplementary information. \href{https://ars.els-cdn.com/content/image/1-s2.0-S0921452624011104-mmc1.pdf}{{\color{blue} Click here to access.}}\\

\noindent \textbf{Acknowledgements}

The authors A.M.M., F.V.P. and G.J. acknowledge Conselho Nacional de Desenvolvimento Científico e Tecnológico (CNPq), Coordena\c{c}\~ao de Aperfei\c{c}oamento Pessoal de Nível Superior (CAPES) and Financiadora de Estudos e Projetos (FINEP) for the financial support. \\

\noindent \textbf{Conflict of Interest} 

The authors declare no conflict of interest.\\

\section*{References}

\bibliography{RefQDots7}
\bibliographystyle{unsrt}

\end{document}